\documentclass[floatfix,aps,prd,showpacs,amsmath,nofootinbib,preprintnumbers,twocolumn]{revtex4}

\usepackage{epsfig}
\usepackage{bm}
\usepackage{amsfonts}

\usepackage{graphicx}

\def\p{\partial}

\def\({\left(}
\def\){\right)}
\def\[{\left[}
\def\]{\right]}
\def\L{\Lambda}

\def\L{\Lambda}
\def\Ol{\Omega_\Lambda}

\def\p{\partial}

\def\p{\partial}

\newcommand{\s}{\sigma}

\newcommand{\vs}{V(\sigma)}

\def\e{\begin{equation}}
\def\q{\end{equation}}
\def\m{\begin{eqnarray}}
\def\n{\end{eqnarray}}

\begin{document}

\title{Theoretical Limits on the Equation-of-State Parameter of Phantom Cosmology}

\author{Emmanuel N.~Saridakis }
\email{msaridak@phys.uoa.gr} \affiliation{Department of Physics,
University of Athens,\\ GR-15771 Athens, Greece}


\begin{abstract}

We investigate the restrictions on the equation-of-state parameter
of phantom cosmology, due to the minimum quantum gravitational
requirements. We find that for all the examined
$w_\Lambda(z)$-parametrizations and for arbitrary phantom
potentials and spatial curvature, the phantom equation-of-state
parameter is not restricted at all. This is in radical contrast
with the quintessence paradigm, and makes phantom cosmology more
robust and capable of constituting the underlying mechanism for
dark energy.

\end{abstract}


\maketitle

\section{Introduction}

Many cosmological observations, such as SNe Ia \cite{1}, WMAP
\cite{2}, SDSS \cite{3} and X-ray \cite{4}, support that the
universe is experiencing an accelerated expansion. These
observations suggest that it is dominated by dark energy with
negative pressure, which provides the dynamical mechanism for such
an accelerating expansion. Although the nature and origin of dark
energy could perhaps understood by a fundamental underlying theory
unknown up to now, physicists can still propose some paradigms to
describe it. In this direction we can consider theories of
modified gravity \cite{ordishov}, or field models of dark energy.
The field models that have been discussed widely in the literature
consider a cosmological constant \cite{cosmo}, a canonical scalar
field (quintessence) \cite{quint}, a phantom field, that is a
scalar field with a negative sign of the kinetic term
\cite{phant,quantumphantom0}, or the combination of quintessence
and phantom in a unified model named quintom \cite{quintom}.
Finally, many theoretical studies are devoted to shed light on
dark energy within the quantum gravitational framework, since,
despite the lack of such a theory at present, we can still make
some attempts to probe the nature of dark energy according to some
of its basic principles. An interesting step in this direction is
the so-called ``holographic dark energy'' proposal \cite{HOLO},
which has been constructed in the light of holographic principle
of quantum gravity  \cite{holoprin}, and thus it presents some
interesting features of an underlying theory of dark energy.

In the present work we are interested in investigating the
theoretical limits on the equation-of-state parameter $w_\L$ of
the phantom paradigm of dark energy, due to the basic requirements
of quantum gravity. As we know, in field dark energy models,
$w_\L$ evolves according to the  field evolution
\cite{phant,quantumphantom0}. Therefore, a basic and necessary
constraint, consistent to our current knowledge of quantum field
theory and (quantum) gravity, should be that the field variation
during cosmological evolution should be less than the Planck mass
$M_p$. Such a constraint on field variation results in limits on
$w_\L$. In the case of quintessence, this investigation has been
performed in \cite{Huang:2007mv} where the corresponding limits
are presented. In this letter we study the phantom scenario.

However, two points must be mentioned here. The first is that a
well-established quantum theory of gravity could possibly induce
stronger limits on $w_\L$. The requirement that the field
variation must be smaller than $M_p$ is just the minimum
condition, consistent with present theoretical knowledge.
Secondly, there is a discussion in the literature whether a
construction of quantum field theory of phantoms is possible,
namely whether the null energy condition is violated
\cite{Hawking:1973uf} leading to causality and stability problems
\cite{Schon:1981vd}. However, more recently there have been
serious attempts in overcoming these difficulties and construct a
phantom theory consistent with the basic requirements of quantum
field theory \cite{quantumphantom0,quantumphantom}. In conclusion,
although the discussion on the aforementioned two points is open,
it is still interesting to examine the limits on $w_\L$ due to the
basic requirement of quantum gravity. The plan of the work is as
follows: In section \ref{phanto} we formulate  the phantom
cosmological scenario and we extract the relation between the
field variation $|\Delta\s|$ and $w_\L(z)$. In section
\ref{cosmimpl} we investigate its behavior for various
$w_\L(z)$-parametrizations and we examine the $w_\L(z)$ limits
implied by the condition $|\Delta\s|<M_p$. Finally, in section
\ref{conclusions} we summarize our results.

\section{Quantum gravitational restrictions on phantom cosmology}
\label{phanto}

 We consider a general Friedmann-Lemaitre-Robertson-Walker
universe with line element
\begin{equation}
\label{metr}
 ds^{2}=-dt^{2}+a^{2}(t)\left(\frac{dr^2}{1-kr^2}+r^2d\Omega^{2}\right)
\end{equation}
 in comoving coordinates  $(t,r,\theta,\varphi)$. $a$ is the scale factor and $k$ denotes
 the spacial curvature, with $k=0,1,-1$ corresponding to a flat, closed
or open universe respectively. The action of a universe
constituted of a phantom  field $\s$ is \cite{phant}:
\begin{eqnarray}
S=M_p^2\int d^{4}x \sqrt{-g} \left[\frac{1}{2} R
+\frac{1}{2}g^{\mu\nu}\p_{\mu}\sigma\p_{\nu}\sigma+\vs
+\cal{L}_\text{M}\right], \label{actionquint}
\end{eqnarray}
where the term $\cal{L}_\text{M}$ accounts for the matter content
of the universe. The Friedmann equations and the evolution
equation for the  phantom field are \cite{phant}:
\begin{equation}
H^{2}=\frac{1}{3M_p^2}\left[\rho_{M}+\rho_{\s}-\rho_k\right],
\label{Fr1}
\end{equation}
\begin{equation}
\left(\frac{\ddot{a}}{a}\right)=-\frac{1}{3M_p^2}\left[\frac{\rho_M}{2}+\frac{3p_M}{2}+2
p_{\s}+V(\s)\right], \label{Fr2}
\end{equation}
\begin{equation}
\ddot{\s}+3H\dot{\s}-\frac{\partial V(\s)}{\partial \s}=0,
\label{phantom}
\end{equation}
where $H=\dot{a}/a$ is the Hubble parameter. In these expressions,
$p_\s$ and $\rho_{\s}$ are respectively the pressure and density
of the phantom field, while $p_M$ and $\rho_M$ are the
corresponding quantities for the matter content of the universe.
Finally, $\rho_k$ stands for the spatial curvature density:
\begin{equation}
\rho_k=3M_p^2\frac{k}{a^2}\label{curvdens}.
\end{equation}

The energy density and pressure of the  phantom field, are given
by:
\begin{equation}
\rho_{\s}=-\frac{1}{2}\dot{\s}^{2}+V(\s)\label{enerphantom}
\end{equation}
\begin{equation}
p_{\s}=-\frac{1}{2}\dot{\s}^{2}-V(\s).\label{enerphantom2}
\end{equation}
As usual, the dark energy of the universe is attributed to the
scalar field and it reads:
\begin{equation}
\Ol=\frac{\rho_\s}{\rho_\s+\rho_M-\rho_k}=\frac{1}{3M_p^2H^2}\left[-\frac{1}{2}\dot{\s}^{2}+V(\s)\right]
\label{OmL}.
\end{equation}
Thus, the equation of state for the  phantom dark energy is
\cite{phant}:
\begin{equation}
w_\L =\frac{p_\s}{\rho_\s}= \frac{-\dot{\s}^{2}-2V(\s)}
{-\dot{\s}^{2}+2V(\s)}\label{eqstate}.
\end{equation}

The equations of motion close by considering the evolution of the
matter density:
\begin{equation}\label{sys3}
\dot{\rho}_M+3H(\rho_M+p_M)=0.
\end{equation}
Finally, we remind that the phantom evolution equation
(\ref{phantom}) can be also written in the form of a conservation
equation, namely:
\begin{equation}\label{sys4}
\dot{\rho}_\s+3H(\rho_\s+p_\s)=0.
\end{equation}

Let us now calculate the phantom field variation in such a general
phantom cosmological scenario. By definition it will be:
\begin{equation}
|\Delta\s|=\int_{\s(z)}^{\s(0)}d\s=\int_{t_z}^{t_0}\dot{\s}d
t=\int_{0}^{z}\dot{\s}\frac{d z'}{H(1+z')}\label{var1},
\end{equation}
where $z$ corresponds to the  redshift of the beginning of the
cosmological evolution (chosen at will), $t_z$ is the
corresponding time, and $z=0$ and $t_0$ are their present values.
The last equality arises from the fact that $Hdt=-\frac{dz}{1+z}$,
according to the standard definition $a=(1+z)^{-1}$, with $a_0=1$
the present value.

The time derivative of the phantom field can be easily calculated
as follows. From the equation-of-state parameter definition
(\ref{eqstate}) we obtain:
\begin{equation}\label{vss}
\vs=\frac{\dot{\s}}{2}\left(\frac{w_\L-1}{w_\L+1}\right).
\end{equation}
Inserting this relation into (\ref{enerphantom}) we acquire:
\begin{equation}\label{vss2}
\dot{\s}=\sqrt{-(w_\L+1)\rho_\s},
\end{equation}
where without loss of generality we have assumed that
$\frac{\partial V(\s)}{\partial \s}<0$, so that $\dot{\s}>0$. Note
that expression  (\ref{vss2}) is always real, since in phantom
scenario $w_\L<-1$ at all times.

Substituting relation (\ref{vss2}) into (\ref{var1}) we obtain:
\begin{eqnarray}
|\Delta\s|&=&\int_{0}^{z}\sqrt{-(w_\L+1)\rho_\s}\left(\frac{\sqrt{3}M_p}{\sqrt{\rho_{M}+\rho_{\s}-\rho_k}}\right)\frac{d
z'}{(1+z')}=\nonumber\\
&=&\int_{0}^{z}\sqrt{3}M_p\sqrt{-[w_\L(z')+1]\Ol(z')}\,\frac{d
z'}{(1+z')} \label{var2},
\end{eqnarray}
where we have also used the Friedmann equation (\ref{Fr1}) and the
definition (\ref{OmL}).

In expression (\ref{var2}), $w_\L$ and $\Ol$ are considered as
functions of $z$. To obtain $\Ol(z)$ we first integrate
(\ref{sys4}), using also the $w_\L$ definition:
\begin{equation}\label{rhos}
\rho_\s(z)=\rho_{\s 0}\,\exp\left[\int_0^z
3\left(\frac{1+w_\L(z')}{ 1+z'}\right)dz'\right].
\end{equation}
In addition, according to (\ref{curvdens}) we have
$\rho_k(z)=\rho_{k0}(1+z)^2$, and as usual
$\rho_M(z)=\rho_{M0}(1+z)^3$. Thus, substituting these relations
for the densities in the $\Ol$ definition (\ref{OmL}) we finally
acquire:
\begin{equation}
\Ol(z)=\left[\frac{\rho_{M}(z)+\rho_{\s}(z)-\rho_k(z)}{\rho_\s(z)}\right]^{-1}=\ \ \ \ \ \ \ \ \ \ \ \ \ \ \ \ \ \ \ \ \ \ \nonumber\\
\end{equation}
{\small{
\begin{equation}
=\left\{1+\left[\frac{\Omega_{M0}}{\Omega_{\L0}}(1+z)-\frac{\Omega_{k0}}{\Omega_{\L0}}\right](1+z)^2\,e^{-\int_0^z
3\left[\frac{1+w_\L(z')}{ 1+z'}\right]dz'}
\right\}^{-1},\label{Olz}
\end{equation}}}
where $\Omega_{M0}$, $\Omega_{\L0}$ and $\Omega_{k0}$ are the
present values of the corresponding density parameters. Therefore,
substituting (\ref{Olz}) into (\ref{var2}) we acquire the desired
 phantom field variation $|\Delta\s|$ as a function of $w_\L(z)$.

As we mentioned in the introduction, the goal of this work is to
investigate the limits on $w_\L(z)$ imposed by the basic
requirements of quantum gravity. In general, in quantum field
theory, in order to calculate the vacuum energy density one has to
sum the zero-point energy of all normal modes of all the fields up
to a UV  cutoff, which is believed to be the Planck mass $M_p$.
However, doing so we result with a vacuum energy tremendously
higher than the observed value. In order to solve this famous
(cosmological constant) problem we have to base upon a quantum
theory of gravity. In \cite{ArkaniHamed:2006dz} it is suggested
that gravity and the other quantum fields  cannot be treated
independently in quantum gravity. For instance, in
four-dimensional Minkowski spacetime a new intrinsic UV cutoff
$gM_p$ arises for the U(1) gauge theory coupled to gravity with
coupling $g$, and this conjecture can be generalized to
asymptotically de Sitter spacetimes \cite{Huang:2006hc}.
Therefore, if there is a U(1) gauge theory with incredibly small
coupling $g\sim 10^{-60}$ in our universe, it will result to a
very small cosmological constant. A similar conjecture can be
proposed in the case of the $\lambda \phi^4$ theory in Minkowski
and asymptotically de Sitter spacetimes \cite{Huang:2007gk}, i.e
that the field value cannot become larger than the Planck scale
$M_p$. Finally, in \cite{Huang:2007qz} this assumption is
generalized to every scalar field model, in order to avoid a
breakdown of the theory due to the transition to over-Planckian
regimes, and this is also supported by string theoretical
arguments \cite{Ooguri:2006in}. In conclusion, in this work we
consider that a minimum and obvious requirement, consistent with
the present knowledge of quantum gravity, is that the phantom
field variation $|\Delta\s|$, throughout the entire cosmological
evolution, must not exceed the Planck mass $M_p$, otherwise it
would have left observable imprints. Thus, using (\ref{var2}), the
condition $|\Delta\s|<M_p$ reads:
\begin{equation}
\frac{|\Delta\s|}{M_p}=
\int_{0}^{z}\sqrt{3}\sqrt{-[w_\L(z')+1]\Ol(z')}\,\frac{d
z'}{(1+z')}<1 \label{var3},
\end{equation}
with $\Ol(z)$ given by (\ref{Olz}).

\section{Theoretical limits on $w_\L(z)$}
\label{cosmimpl}

In the previous section we extracted the minimum quantum
gravitational restriction on phantom cosmology, namely relation
(\ref{var3}). Our strategy is to use various parametrizations of
$w_\L(z)$ (since there is not a single, fundamental parametric
form \cite{Szydlowski:2006ay}) in order to extract the
restrictions on their parameters according to (\ref{var3}).
Finally, we will use the standard values $\Omega_{\L0}=0.73$ and
$\Omega_{M0}=0.27$. Concerning the curvature density
$\Omega_{k0}$, we will assume it to be zero, as motivated by
theoretical considerations, such as inflation, and observations.
We will discuss the $\Omega_{k0}\neq0$ scenarios in the end of
this section. Let us
now investigate the various $w_\L(z)$ cases of the literature.\\

\begin{center}Case I: $w_\L(z)=w_0=$const \end{center}

\begin{figure}[ht]
\begin{center}
\mbox{\epsfig{figure=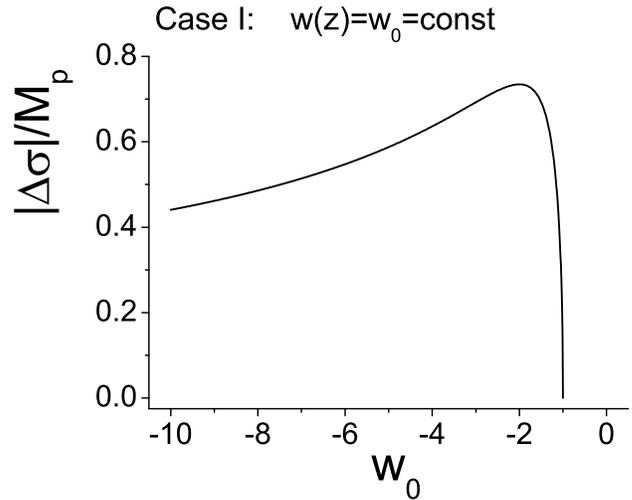,width=8.3cm,angle=0}} \caption{{\it
$|\Delta\s|/M_p$ according to (\ref{var3}), for
$w_\L(z)=w_0=$const (Case I) and $z=z_{rec}=1089$. }}
\label{caseI}
\end{center}
\end{figure}

We start our study by the simplest model, that is a constant
$w_\L<-1$. As an ``initial'' $z$ for the cosmological evolution we
will consider the last scattering, that is $z=z_{rec}=1089$,
however our quantitative results are almost independent of $z$ for
$z>2$. In fig.~\ref{caseI} we depict $|\Delta\s|/M_p$ according to
(\ref{var3}), for $w_\L(z)=w_0=$const.
Surprisingly enough, we observe that for every value of the
parameter $w_0$, the ratio $|\Delta\s|/M_p$ is always less than
one. Thus, the minimum quantum gravitational restriction does not
imply any limit on $w_0$ in phantom cosmology. This is in radical
contrast with the corresponding result for quintessence paradigm,
where for this simple $w_\L(z)$-parametrization the author finds
$w=w_0\leq -0.738$ \cite{Huang:2007mv}.

\begin{center}Case II. $w_\L(z)=w_0+w_1z$\end{center}

\begin{figure}[ht]
\begin{center}
\mbox{\epsfig{figure=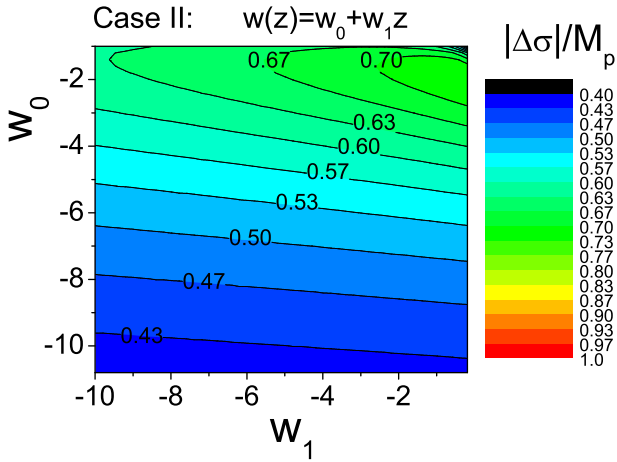,width=8.3cm,angle=0}}
\caption{(Color Online) {\it $|\Delta\s|/M_p$ according to
(\ref{var3}), for $w_\L(z)=w_0+w_1z$ (Case II) and $z=2$. }}
\label{caseII}
\end{center}
\end{figure}

Let us consider the case where the equation-of-state parameter is
a linear function of the redshift \cite{Cooray:1999da}. This
proves to be a good parametrization at low redshift, in agreement
with observations. However, $w_\L(z)$ obviously diverges at large
$z$, making it unsuitable at high redshift. As we know, the
redshift of the Supernova Legacy Survey  is less than 2 \cite{1},
and thus we will use this value as an ``initial'' redshift of the
phantom cosmological evolution. In fig.~\ref{caseII} we depict
$|\Delta\s|/M_p$ according to (\ref{var3}), for $w_\L(z)=w_0+w_1z$
and $z=2$.
As we observe,  $|\Delta\s|/M_p$ is always less than one,
independently of the values of the parameters $w_0$ and $w_1$.
This result holds even if we consider another term in the
parametrization, namely $w_\L(z)=w_0+w_1z+w_2z^2$ (following
\cite{Szydlowski:2008zzaB}), and even if we consider another value
for the ``initial'' $z$ (we mention that our quantitative results
are independent of $z$ for $z>4$). Thus, the condition
$|\Delta\s|<M_p$ does not imply any limit on the parameters of
this $w_\L(z)$-parametrization in phantom cosmology.

\begin{figure}[ht]
\begin{center}
\mbox{\epsfig{figure=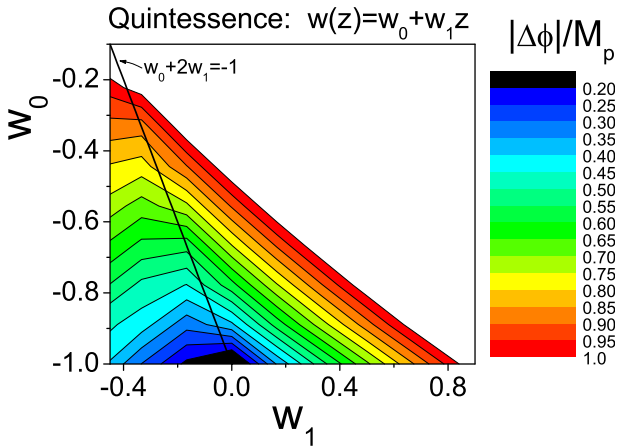,width=8.3cm,angle=0}}
\caption{(Color Online) {\it $|\Delta\phi|/M_p$ for
$w_\L(z)=w_0+w_1z$ (Case II) and $z=2$ in the case of quintessence
cosmology. The straight line marks the $w_\L(z=2)=-1$ region, thus
only the area on the right of this line is physically meaningful
for the quintessence scenario.}} \label{caseIIQ}
\end{center}
\end{figure}

Again, this is in  contrast with the corresponding result for
quintessence paradigm, where it can be shown that  $-1\leq w_0\leq
-0.204$ and $-0.417\leq w_1\leq 0.854$ \cite{Huang:2007mv}. To
present this qualitative difference in a more transparent way, we
repeat our investigation for this  $w_\L(z)$-parametrization, but
for a canonical $\phi$ instead of a phantom field, i.e for the
case of quintessence. In this case we result to a relation similar
to (\ref{var3}), but without the minus sign in the square root. In
fig.~\ref{caseIIQ} we depict $|\Delta\phi|/M_p$ for
$w_\L(z)=w_0+w_1z$ in the case of quintessence cosmology.
Clearly, the constraint $|\Delta\phi|<M_p$ leads to the
aforementioned limits on $w_0$ and $w_1$.

\begin{center}Case III. $w_\L(z)=w_0+w_1\frac{z}{1+z}$ \end{center}

\begin{figure}[ht]
\begin{center}
\mbox{\epsfig{figure=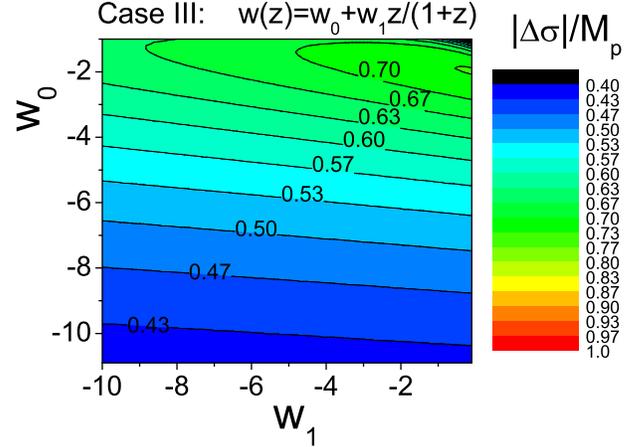,width=8.3cm,angle=0}}
\caption{(Color Online) {\it $|\Delta\s|/M_p$ according to
(\ref{var3}), for $w_\L(z)=w_0+w_1\frac{z}{1+z}$ (Case III) and
$z=z_{rec}=1089$. }} \label{caseIII}
\end{center}
\end{figure}

This parametrization is suggested in \cite{Chevallier:2000qy}. It
overcomes the divergence problem of case II above and has been
widely used in the literature. In fig.~\ref{caseIII} we depict
$|\Delta\s|/M_p$ according to (\ref{var3}), for
$w_\L(z)=w_0+w_1\frac{z}{1+z}$ and $z=z_{rec}=1089$.
Similarly to the previous cases, we see that $|\Delta\s|/M_p$ is
always less than one, independently of the values of the
parameters $w_0$ and $w_1$. This holds even if we consider another
term in the parametrization, namely
$w_\L(z)=w_0+w_1\frac{z}{1+z}+w_2\left(\frac{z}{1+z}\right)^2$,
following \cite{Szydlowski:2008zzaB}. Finally, these results hold
even if we consider another value for the ``initial'' $z$ (we
mention that in this case our quantitative results are independent
of $z$ for $z>10$). Therefore, the condition $|\Delta\s|<M_p$ does
not imply any limit on the parameters of this
$w_\L(z)$-parametrization in phantom cosmology. This is also in
contrast with the corresponding quintessence case, where we obtain
$-1\leq w_0\leq -0.434$ and $-0.564\leq w_1\leq 0.498$
\cite{Huang:2007mv}.

\begin{center}Case IV. $w_\L(z)=w_0+w_1\ln(1+z)$\end{center}

\begin{figure}[ht]
\begin{center}
\mbox{\epsfig{figure=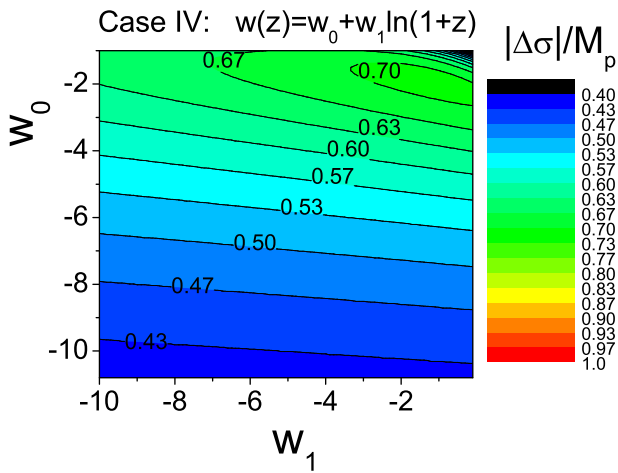,width=8.3cm,angle=0}}
\caption{(Color Online) {\it $|\Delta\s|/M_p$ according to
(\ref{var3}), for $w_\L(z)=w_0+w_1\ln(1+z)$ (Case IV) and $z=2$.
}} \label{caseIV}
\end{center}
\end{figure}

\begin{figure}[ht]
\begin{center}
\mbox{\epsfig{figure=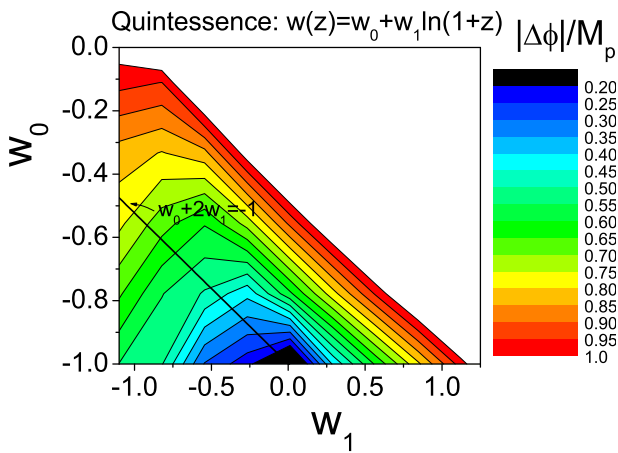,width=8.3cm,angle=0}}
\caption{(Color Online) {\it $|\Delta\phi|/M_p$ for
$w_\L(z)=w_0+w_1\ln(1+z)$ (Case IV) and $z=2$ in the case of
quintessence cosmology. The straight line marks the $w_\L(z=2)=-1$
region, thus only the area on the right of this line is physically
meaningful for the quintessence scenario.}} \label{caseIVQ}
\end{center}
\end{figure}

This parametrization is suggested in \cite{Szydlowski:2008zzaB}.
At low redshift it is very efficient in describing observations,
but at high redshift it diverges, although more slowly than case
II above. Thus, as an ``initial'' redshift we will consider $z=2$.
In fig.~\ref{caseIV} we present $|\Delta\s|/M_p$ according to
(\ref{var3}), for $w_\L(z)=w_0+w_1\ln(1+z)$.
We observe that $|\Delta\s|/M_p$ is always less than one,
independently of the values of the parameters $w_0$ and $w_1$.
This result holds even if we consider a third term in the
parametrization, namely
$w_\L(z)=w_0+w_1\ln(1+z)+w_2\left[\ln(1+z)\right]^2$ (as suggested
in \cite{Szydlowski:2008zzaB}). Furthermore, these results hold
even if we consider another value for the ``initial'' $z$, since
quantitatively our results are independent of $z$ for $z>6$. Thus,
the condition $|\Delta\s|<M_p$ does not imply any limit on the
parameters of this $w_\L(z)$-parametrization in phantom scenario.
This is in contrast with the corresponding quintessence case,
which we present in fig.~\ref{caseIVQ} since this case has not
been studied in the literature.
In this scenario we get $-1\leq w_0\leq -0.061$ and $-1.101\leq
w_1\leq 1.159$. The qualitative difference between phantom and
quintessence behavior is obvious.\\

\begin{center}Case V. Phantom Models with Nearly Flat Potentials\end{center}

\begin{figure}[ht]
\begin{center}
\mbox{\epsfig{figure=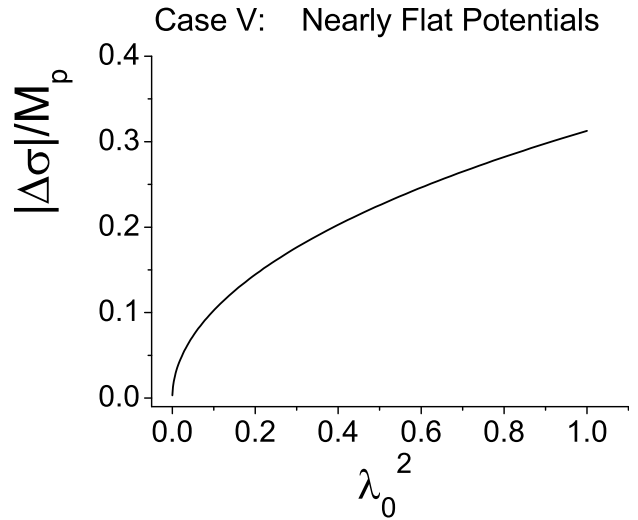,width=8.3cm,angle=0}}
\caption{{\it $|\Delta\s|/M_p$ for phantom models with nearly flat
potentials (Case V), with $w_\L(z)$ given by (\ref{phfl}) and
$z=z_{rec}=1089$. }} \label{caseV}
\end{center}
\end{figure}

In \cite{phantflat} the authors examine phantom models with nearly
flat potentials. Under this assumption they result in a single
expression for $w_\Lambda(z)$, depending only on the initial field
values and their derivatives. In particular, they obtain the
following $w_\L(z)$-parametrization:
 {\small{
\begin{equation}
w_\L(z) =-1 - \ \ \ \ \ \ \ \ \ \ \ \ \ \ \ \ \ \ \ \  \ \ \ \ \ \
\ \ \ \ \ \ \ \ \ \ \ \ \ \  \ \ \ \ \ \ \ \ \ \ \ \ \ \ \ \ \ \ \
\ \nonumber
 \end{equation}
\begin{equation}
-\frac{\lambda_0^2}{3}\left[\frac{1}{\sqrt{\Ol(z)}} -
\frac{1}{2}\left(\frac{1}{\Ol(z)} - 1 \right) \ln
\left(\frac{1+\sqrt{\Ol(z)}} {1-\sqrt{\Ol(z)}}
\right)\right]^2\label{phfl},
 \end{equation}}}
with the single parameter $\lambda_0$ satisfying $\lambda_0\ll1$.
In fig.~\ref{caseV} we present $|\Delta\s|/M_p$ for this case, and
since $w_\L(z)$ does not diverge for high $z$ we use
$z=z_{rec}=1089$ (note that our quantitative results do not depend
on $z$ for $z>3$).
As we can see $|\Delta\s|/M_p$ is always less than one,
independently of the value of the parameter $\lambda_0$. Thus, the
condition $|\Delta\s|<M_p$ does not imply any limit on the
parameters of this $w_\L(z)$-parametrization in phantom scenario.

In the investigation of this section, up to now, we have
considered $\Omega_{k0}=0$, that is a flat universe. Let us now
examine the $\Omega_{k0}\neq0$ case. Repeating the same steps we
find that the requirement $|\Delta\s|<M_p$ is satisfied without
implying any restrictions on $w_\L(z)$, as long as
$-1<\Omega_{k0}<0.25$. Although the value of $\Omega_{k0}$ cannot
be determined exactly by observations, it is highly unlikely to
exceed this range. Thus, our results are valid also in the
non-flat scenarios, providing $\Omega_{k0}$ takes realistic
values.

\section{Conclusions}
\label{conclusions}

In this work we investigate the possible limitations on the
phantom equation-of-state parameter $w_\L(z)$, due to quantum
gravitational effects. Since quantum gravity is still a matter of
research, we have to rely on the basic and minimum requirement of
our current knowledge on the field, namely that the field
variation during the entire cosmological evolution must not exceed
the Planck scale, otherwise it would have left observable
imprints. Although a well-established quantum theory of gravity
could possibly induce stronger limits on $w_\L$, it is still
interesting to investigate the aforementioned condition. Finally,
note that a restriction based on the field values is more general
and more fundamental than one based on the potential (for example
$|V/V'|<M_p$ as considered in in \cite{Caldwell:2005tm}).

Surprisingly enough, we find that for various
$w_\L(z)$-parametrizations, the condition $|\Delta\s|<M_p$ in
phantom cosmology does not imply any limitations on $w_\L(z)$ at
all. This is in radical contrast with the quintessence case, where
even this minimum requirement results in strong limitations on
$w_\L(z)$, even more stringent than the present experiments
\cite{2,Huang:2007mv}. The reason behind this difference is the
sign change in some of the corresponding expressions of the two
cosmological scenarios, as well as the fact that $w(z)<-1$ in
phantom while $w(z)>-1$ in quintessence models. These features
lead the phantom quantities to behave more smoothly, comparing to
the quintessence ones, and thus the simple quantum gravitational
condition is not violated.

In our investigation the phantom potential can be arbitrary, and
thus our results are general and hold for every phantom
cosmological scenario. Furthermore, they are valid in the non-flat
universe, too. In conclusion, we see that the phantom paradigm,
and its induced dark energy equation-of-state parameter, is not at
all restricted by the basic quantum gravitational requirement.
This feature makes phantom cosmology more robust and capable of
composing the
underlying mechanism for dark energy.\\

\paragraph*{{\bf{Acknowledgements:}}}
The author wishes to thank Institut de Physique Th\'eorique, CEA,
for the hospitality during the preparation of the present work.


\begin{thebibliography}{99}



\bibitem{1}
  A.~G.~Riess {\it et al.}  [Supernova Search Team Collaboration],
  Astron.\ J.\  {\bf 116}, 1009 (1998);
S. Perlmutter {\it{et al.}} [Supernova Cosmology Project
Collaboration], Astrophys. J. {\bf 517}, 565 (1999).

\bibitem{2}
C. L. Bennett {\it{et al.}}, Astrophys. J. Suppl. {\bf 148}, 1
(2003); D.~N.~Spergel {\it et al.}  [WMAP Collaboration],
Astrophys.\ J.\ Suppl.\  {\bf 170}, 377 (2007).

\bibitem{3}
M. Tegmark {\it{et al.}} [SDSS Collaboration], Phys. Rev. D {\bf
69}, 103501 (2004).

\bibitem{4}
S. W. Allen, {\it{et al.}}, Mon. Not. Roy. Astron. Soc. {\bf 353},
457 (2004).

\bibitem{ordishov}
S.Nojiri and S.~D.~Odintsov, Phys. Rev. D {\bf{68}}, 123512
(2003);
  P.~S.~Apostolopoulos, N.~Brouzakis, E.~N.~Saridakis and N.~Tetradis,
  Phys.\ Rev.\  D {\bf 72}, 044013 (2005);
 S.Nojiri and S.~D.~Odintsov, Int. J. Geom. Meth. Mod. Phys.
{\bf{4}}, 115 (2007);
  M.~R.~Setare and E.~N.~Saridakis,
  Phys.\ Lett.\  B {\bf 670}, 1 (2008).





\bibitem{cosmo}  P. J. Peebles and B. Ratra,  Rev. Mod. Phys. {\bf{ 75}},
559 (2003);  J. Kratochvil, A. Linde, E. V. Linder and M.
Shmakova, JCAP  {\bf{0407}} 001 (2004);
   F.~K.~Diakonos and E.~N.~Saridakis,
  JCAP {\bf 0902}, 030 (2009).



\bibitem{quint} R. R. Caldwell, R. Dave and P. J. Steinhardt, Phys. Rev. Lett. {\bf{80}},
1582 (1998);  M. S. Turner and M. White, Phys. Rev. D {\bf{56}},
4439 (1997); T. Chiba, Phys. Rev. D {\bf{60}}, 083508 (1999);
Z.~K.~Guo, N.~Ohta and Y.~Z.~Zhang, Phys.\ Rev.\  D {\bf 72},
023504 (2005); Z.~K.~Guo, N.~Ohta and Y.~Z.~Zhang, Mod.\ Phys.\
Lett.\  A {\bf 22}, 883 (2007).

\bibitem{phant} R. R. Caldwell, Phys.
Lett. B {\bf{545}}, 23 (2002); R.~R.~Caldwell, M.~Kamionkowski and
N.~N.~Weinberg, Phys. Rev. Lett. {\bf 91}, 071301 (2003); S.
Nojiri and S. D. Odintsov, Phys. Rev. D  {\bf{72}}, 023003 (2005);
H. Garcia-Compean, G. Garcia-Jimenez,  O. Obregon, and C. Ramirez,
JCAP 0807, 016 (2008);
  M.~Jamil, M.~A.~Rashid and A.~Qadir,
  Eur.\ Phys.\ J.\  C {\bf 58}, 325 (2008);
  M.~R.~Setare and E.~N.~Saridakis,
  JCAP {\bf 0903}, 002 (2009);
  X.~m.~Chen, Y.~g.~Gong and E.~N.~Saridakis,
  JCAP {\bf 0904}, 001 (2009).


\bibitem{quantumphantom0}
  S.~Nojiri and S.~D.~Odintsov,
  Phys.\ Lett.\  B {\bf 562}, 147 (2003)
  [arXiv:hep-th/0303117].

\bibitem{quintom}
B.~Feng, X.~L.~Wang and X.~M.~Zhang, Phys.\ Lett.\  B {\bf 607},
35 (2005);
Z. K. Guo, {\it{et al.}}, Phys. Lett. B {\bf 608}, 177 (2005);
M.-Z Li, B. Feng, X.-M Zhang, JCAP, 0512, 002 (2005); B. Feng, M.
Li, Y.-S. Piao and X. Zhang, Phys. Lett. B {\bf 634}, 101 (2006);
M. R. Setare, Phys. Lett. B {\bf 641}, 130 (2006); W. Zhao and Y.
Zhang, Phys. Rev. D {\bf73}, 123509, (2006);
 M. R.
Setare, J. Sadeghi, and A. R. Amani, Phys. Lett. B {\bf 660}, 299
(2008); J. Sadeghi, M. R. Setare, A. Banijamali and F. Milani,
Phys. Lett. B {\bf 662}, 92 (2008);
  M.~R.~Setare and E.~N.~Saridakis,
  Phys.\ Lett.\  B {\bf 668}, 177 (2008);
  M.~R.~Setare and E.~N.~Saridakis,
  [arXiv:0807.3807 [hep-th]];
  M.~R.~Setare and E.~N.~Saridakis,
  JCAP {\bf 0809}, 026 (2008).


\bibitem{HOLO}
  A.~G.~Cohen, D.~B.~Kaplan and A.~E.~Nelson,
  Phys.\ Rev.\ Lett.\  {\bf 82}, 4971 (1999);
  P.~Horava and D.~Minic,
  Phys.\ Rev.\ Lett.\  {\bf 85}, 1610 (2000);
  S.~D.~H.~Hsu,
  Phys.\ Lett.\ B {\bf 594}, 13 (2004);
  M.~Li,
  Phys.\ Lett.\ B {\bf 603}, 1 (2004);
  D.~Pavon and W.~Zimdahl,
  Phys.\ Lett.\ B {\bf 628}, 206 (2005);
    M.~R.~Setare,
  Phys.\ Lett.\ B {\bf 642}, 1 (2006);
  H.~Li, Z.~K.~Guo and Y.~Z.~Zhang,
  Int.\ J.\ Mod.\ Phys.\ D {\bf 15}, 869 (2006);
M. R. Setare, J. Zhang and X. Zhang, JCAP {\bf 0703}, 007 (2007);
M. R. Setare, Phys. Lett. B {\bf 654}, 1 (2007); W. Zhao, Phys.
Lett. B {\bf 655}, 97 (2007); M. Li, C. Lin and Y. Wang, JCAP {\bf
0805}, 023 (2008);
  E.~N.~Saridakis,
  Phys.\ Lett.\  B {\bf 660}, 138 (2008);
  E.~N.~Saridakis,
  JCAP {\bf 0804}, 020 (2008);
  E.~N.~Saridakis,
  Phys.\ Lett.\  B {\bf 661}, 335 (2008).


\bibitem{holoprin}
G.~'t Hooft,
  [arXiv:gr-qc/9310026];
 L.~Susskind,
  J.\ Math.\ Phys.\  {\bf 36}, 6377 (1995).

\bibitem{Huang:2007mv}
  Q.~G.~Huang,
  Phys.\ Rev.\  D {\bf 77}, 103518 (2008).

\bibitem{Hawking:1973uf}
  S.~W.~Hawking and G.~F.~R.~Ellis,
  ``The Large scale structure of space-time,''
{\it  Cambridge University Press, Cambridge, 1973}.

\bibitem{Schon:1981vd}
  R.~Schon and S.~T.~Yau,
  Commun.\ Math.\ Phys.\  {\bf 79}, 231 (1981);
  E.~Witten,
  Commun.\ Math.\ Phys.\  {\bf 80}, 381 (1981).


\bibitem{quantumphantom}
  S.~Nojiri and S.~D.~Odintsov,
  Phys.\ Lett.\  B {\bf 571}, 1 (2003);
  D.~Samart and B.~Gumjudpai,
  Phys.\ Rev.\  D {\bf 76}, 043514 (2007).

\bibitem{ArkaniHamed:2006dz}
  N.~Arkani-Hamed, L.~Motl, A.~Nicolis and C.~Vafa,
  JHEP {\bf 0706}, 060 (2007).

\bibitem{Huang:2006hc}
  Q.~G.~Huang, M.~Li and W.~Song,
  JHEP {\bf 0610}, 059 (2006).

\bibitem{Huang:2007gk}
  Q.~G.~Huang,
  JHEP {\bf 0705}, 096 (2007).

\bibitem{Huang:2007qz}
  Q.~G.~Huang,
  Phys.\ Rev.\  D {\bf 76}, 061303 (2007).


\bibitem{Ooguri:2006in}
  H.~Ooguri and C.~Vafa,
  Nucl.\ Phys.\  B {\bf 766}, 21 (2007).

\bibitem{Szydlowski:2006ay}
  M.~Szydlowski, A.~Kurek and A.~Krawiec,
  Phys.\ Lett.\  B {\bf 642}, 171 (2006);
  E.~J.~Copeland, M.~Sami and S.~Tsujikawa,
  Int.\ J.\ Mod.\ Phys.\  D {\bf 15}, 1753 (2006);
  H.~Wei and S.~N.~Zhang,
  Phys.\ Lett.\  B {\bf 644}, 7 (2007).

\bibitem{Cooray:1999da}
  A.~R.~Cooray and D.~Huterer,
  Astrophys.\ J.\  {\bf 513}, L95 (1999);
  E.~Di Pietro and J.~F.~Claeskens,
  Mon.\ Not.\ Roy.\ Astron.\ Soc.\  {\bf 341}, 1299 (2003).

\bibitem{Szydlowski:2008zzaB}
  M.~Szydlowski, O.~Hrycyna and A.~Kurek,
  Phys.\ Rev.\  D {\bf 77}, 027302 (2008).

\bibitem{Chevallier:2000qy}
  M.~Chevallier and D.~Polarski,
  Int.\ J.\ Mod.\ Phys.\  D {\bf 10}, 213 (2001);
  E.~V.~Linder,
  Phys.\ Rev.\ Lett.\  {\bf 90}, 091301 (2003).


\bibitem{phantflat}
  R.~J.~Scherrer and A.~A.~Sen,
  Phys.\ Rev.\  D {\bf 78}, 067303 (2008)
  [arXiv:0808.1880[astro-ph]];
  M.~R.~Setare and E.~N.~Saridakis,
   Phys.\ Rev.\  D {\bf 79}, 043005 (2009)
  [arXiv:0810.4775[astro-ph]].

\bibitem{Caldwell:2005tm}
  R.~R.~Caldwell and E.~V.~Linder,
  Phys.\ Rev.\ Lett.\  {\bf 95}, 141301 (2005)
  [arXiv:astro-ph/0505494].





\end{thebibliography}
\end{document}